\documentclass[aps,prl,superscriptaddress,floatfix,reprint,showkeys,showpacs]{revtex4-1}

\usepackage{graphicx}
\usepackage{epstopdf}

\usepackage{amsmath,amssymb,amstext}
\usepackage{hyperref}

\newif\ifshowsupplemental
\showsupplementaltrue

\begin{document}
  \title{Imaging Optical Frequencies with 100 $\mu$Hz Precision and 1.1 $\mu$m Resolution} 
\author{G. Edward Marti}
\email{edward.marti@jila.colorado.edu}
\affiliation{JILA, National Institute of Standards and Technology and University of Colorado, 440 UCB, Boulder, CO 80309, USA}
\affiliation{Department of Physics, University of Colorado, 390 UCB, Boulder, CO 80309, USA}

\author{Ross B. Hutson}
\affiliation{JILA, National Institute of Standards and Technology and University of Colorado, 440 UCB, Boulder, CO 80309, USA}
\affiliation{Department of Physics, University of Colorado, 390 UCB, Boulder, CO 80309, USA}

\author{Akihisa Goban}
\affiliation{JILA, National Institute of Standards and Technology and University of Colorado, 440 UCB, Boulder, CO 80309, USA}
\affiliation{Department of Physics, University of Colorado, 390 UCB, Boulder, CO 80309, USA}

\author{Sara L. Campbell}
\affiliation{JILA, National Institute of Standards and Technology and University of Colorado, 440 UCB, Boulder, CO 80309, USA}
\affiliation{Department of Physics, University of Colorado, 390 UCB, Boulder, CO 80309, USA}

\author{Nicola Poli}
\affiliation{Dipartimento di Fisica e Astronomia and LENS - Universit\`a di Firenze, INFN - Sezione di Firenze, Via Sansone 1, 50019 Sesto Fiorentino, Italy}
\affiliation{Istituto Nazionale di Ottica, Consiglio Nazionale delle Ricerche (INO-CNR), Largo Enrico Fermi, 6, 50125 - Firenze, Italy}

\author{Jun Ye}
\affiliation{JILA, National Institute of Standards and Technology and University of Colorado, 440 UCB, Boulder, CO 80309, USA}
\affiliation{Department of Physics, University of Colorado, 390 UCB, Boulder, CO 80309, USA}

\date{November 22, 2017}
\begin{abstract}
We implement imaging spectroscopy of the optical clock transition of lattice-trapped degenerate fermionic Sr in the Mott-insulating regime, combining micron spatial resolution  with submillihertz spectral precision. We use these tools to demonstrate atomic coherence for up to 15 s on the clock transition and reach a record frequency precision of $2.5\times 10^{-19}$. We perform the most rapid evaluation of trapping light shifts and record a 150 mHz linewidth, the narrowest Rabi line shape observed on a coherent optical transition. The important emerging capability of combining high-resolution imaging and spectroscopy will improve the clock precision, and provide a path towards measuring many-body interactions and testing fundamental physics.
 \end{abstract}
\maketitle
Alkaline-earth (AE) atoms possess ultranarrow optical (``clock'') transitions that realize the best atomic clocks \cite{Bloom2014,Ushijima2015,Nicholson2015,Schioppo2017,Campbell2017}.
AE atoms become sensitive probes of their external environment, of interactions, and of fundamental physics through highly precise measurements of the optical clock frequency.
Recently, frequency shifts that arise from atomic interactions in ultracold samples have been used to study magnetism and spin-orbit coupling in nondegenerate ensembles of AE atoms \cite{Martin2013a, Zhang2014, Kolkowitz2017}; as well as spin-exchange processes, Feshbach resonances, and synthetic dimensions in degenerate samples \cite{Scazza2014, Cappellini2014, Hofer2015, Mancini2015, Cappellini2014}.

Combining \textit{in situ} imaging with state-of-the-art optical spectroscopy provides a new route to improve the precision of atomic clocks, study both few- and many-body phenomena, and test fundamental physics. 
Imaging the relative clock frequency between atoms in different regions of the optical lattice, called imaging spectroscopy, allows for synchronous frequency comparisons that improve precision by rejecting laser frequency noise and common-mode clock shifts.
In particular, frequency differences can be measured at the quantum-projection-noise (QPN) limit by comparing the Ramsey spectroscopy excitation fraction of one atomic ensemble against another, even when the free-evolution time exceeds the laser coherence time.
In addition to determining single-particle effects such as lattice light shifts that impact optical clock accuracy, maps of the local frequencies can elucidate few-body physics of atoms interacting within a lattice site, and many-body interactions between lattice sites.
Thus, imaging spectroscopy gives information analogous to scanning tunneling microscopy, and will enable the exploration of long-range electric dipole-dipole interactions \cite{Chang2004a, Kraemer2016} and new phenomena, such as the Kondo effect \cite{Foss-Feig2010, Riegger2017}, SU($N$) quantum magnetism \cite{Honerkamp2004,Hermele2009,Gorshkov2010}, and unconventional superconductivity \cite{Hofstetter2002,Trebst2006,Rey2009}.
Since synchronous comparisons improve precision and hence allow for more rapid measurements than conventional techniques, they bring new tests of gravitational and other fundamental physics within the range of tabletop experiments. At $10^{-19}$ fractional frequency precision, gravitational redshifts can be measured within a single vacuum chamber, opening the door to exploring the interplay of quantum mechanics and general relativity \cite{Chou2010a, Zych2011}.

\begin{figure}[htb!]
\begin{center}
\includegraphics[scale=1.0]{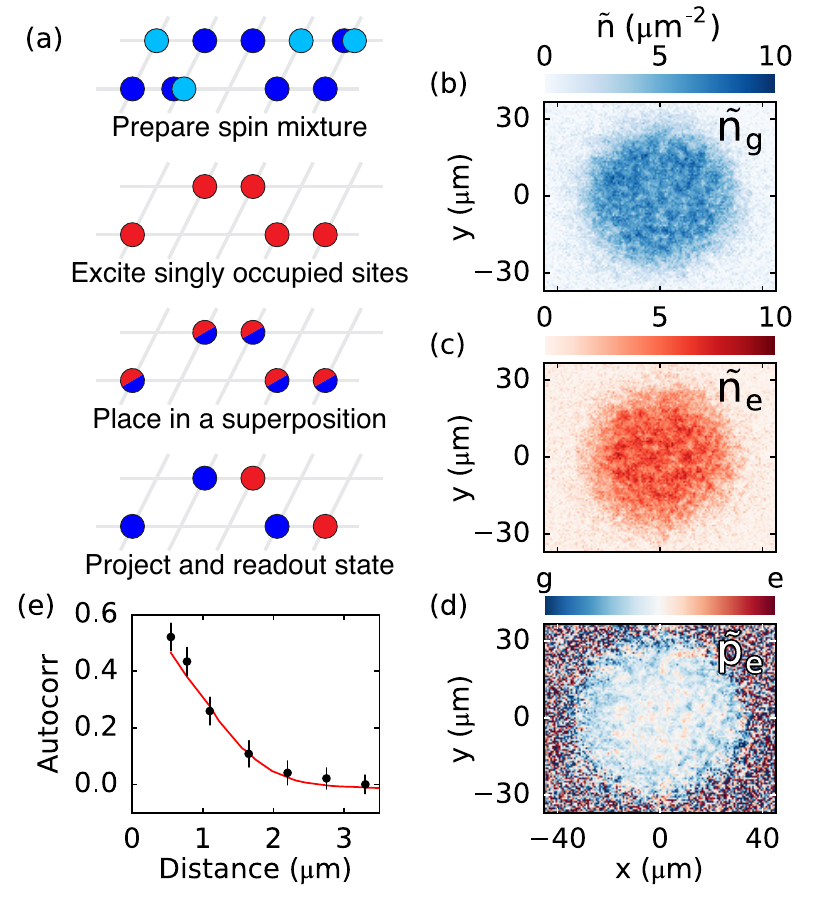}
\caption{Imaging spectroscopy experimental sequence.
    (a) A spin mixture of $|{}^1\mathrm{S}_0, m_F=\pm 1/2\rangle$ atoms is loaded into a 3D optical lattice.
    A $\pi$ pulse drives singly occupied sites of one spin state to the excited state, after which all other atoms are removed.
    The remaining atoms are placed in a superposition state and then read out with absorption imaging.
    Images are processed to yield the column density of (b) the ground state, $\tilde n_g$, (c) the excited state, $\tilde n_e$, and (d) the excitation fraction $\tilde p_e$.
    Small spatial frequency shifts can be measured through changes in $\tilde p_e$.
    (e) The normalized autocorrelation function of the central $15\;\mu\mathrm{m} \times 10\;\mu\mathrm{m}$ region  of $\tilde p_e$ (black points) demonstrates correlations induced by the imaging system that correspond to a $1.1\;\mu\mathrm{m}$ imaging resolution (red line).
\label{fig:schematic}}
\end{center}
\end{figure}

\begin{figure*}[t!]
\begin{center}
\includegraphics[scale=0.98]{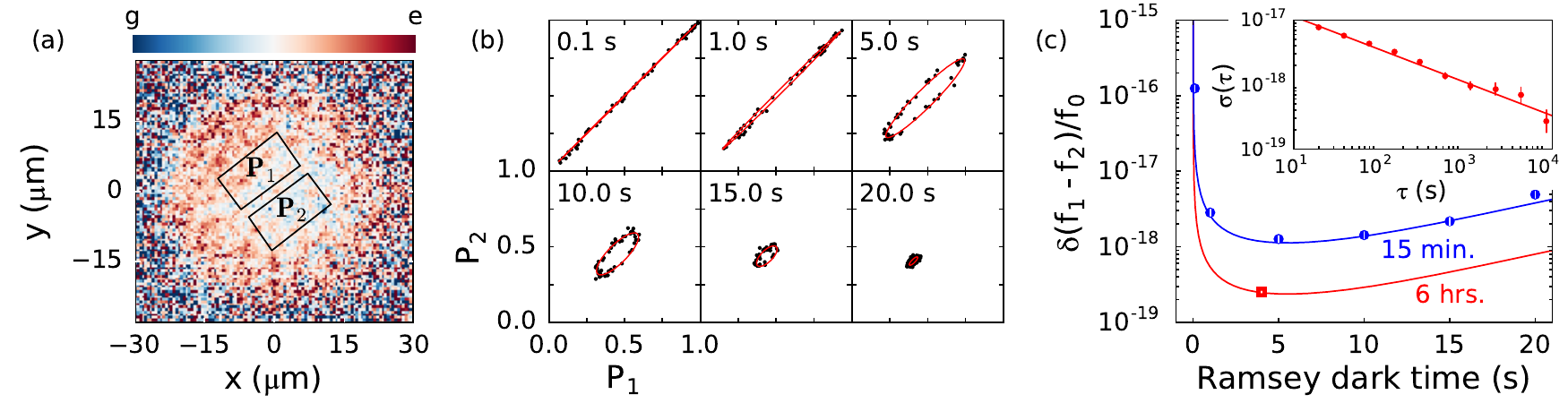}
\caption{Imaging the local excitation fraction allows for the determination of small frequency shifts within the lattice-trapped sample.
    (a) A map of the excitation fraction of atoms after a Ramsey sequence ($T=5$ s) is split into two separate regions.
    (b) Parametric plots of $P_1$ against $P_2$ (black points)  show ellipses, created by a reproducible phase shift between the two regions.
    A maximum likelihood estimator determines the ellipse properties (red line). 
    The phase shift increases with $T$ for a fixed frequency difference $f_1 - f_2$, while the contrast decays. 
    (c) The measured uncertainty $\delta(f_1 - f_2)$ for 900 s averaging time (blue points) closely follows the expected QPN limit for ellipse fitting (blue line).
    The measurements remain QPN-limited for 1,000 experimental repetitions (red square and line), reaching a fractional uncertainty of $2.5\times 10^{-19}$, with (inset) a total Allan deviation that averages with a slope of $3.6 \times 10^{-17}/\sqrt{\mathrm{Hz}}$.
    This uncertainty would correspond to a 2.3 mm gravitational redshift on the Earth.
    The frequencies are normalized to the clock frequency $f_0\approx 429$ THz of $^{87}$Sr. 
\label{fig:signal_to_noise}}
\end{center}
\end{figure*}

In this Letter, we perform imaging spectroscopy on a two-state spin mixture of Fermi-degenerate ${}^{87}$Sr prepared in the Mott-insulating regime of a three-dimensional (3D) optical lattice with submillihertz-precision optical spectroscopy and micron-resolution spatial imaging.
This work builds on the long coherence time demonstrated in Ref.~\onlinecite{Campbell2017} and leverages high-resolution imaging for the interrogation of strontium atoms in a 3D optical lattice clock.
First, we characterize the resolution of the imaging system.
We then demonstrate a QPN-limited frequency difference measurement between two regions of the lattice. Using thousands of atoms that remain coherent for up to 15 s, we reach a record in frequency precision of $2.5\times 10^{-19}$, or 100 $\mu$Hz on an optical frequency.
This excellent precision allows us to measure the shift of the clock transition by the optical lattice with a spatially-dependent frequency map.
Imaging spectroscopy provides a clear path towards reducing the uncertainty of the optical lattice light shift by more than an order of magnitude.
Finally, we use imaging spectroscopy as a multiplexed measurement of the frequency noise of an ultrastable laser.
This is accomplished with a magnetic field gradient such that different regions of the lattice simultaneously probe different components of the laser frequency.
The lattice then acts as a highly multiplexed optical spectrum analyzer.
Similarly, the lattice can be employed as a multiaxis sensor of electromagnetic, gravitational, or other field gradients.

Atomic preparation follows Ref.~\onlinecite{Campbell2017}. In summary, nearly $10^7$ ${}^{87}$Sr atoms are laser-cooled to 3 $\mu$K in a crossed optical dipole trap.
Atoms are optically pumped to an incoherent mixture of the two $|{}^1 \mathrm{S}_0, m_F=\pm 1/2\rangle$ states.
Forced evaporative cooling lowers the temperature to $15\,\mathrm{nK}=0.1\,T_F$, where $T_F$ is the Fermi temperature, with $10^4$ atoms per spin state.
The two-spin-state mixture is then adiabatically loaded into a deep 3D optical lattice with typical trap frequencies of $2\pi \times 50$ kHz and negligible tunneling.
A homogenous magnetic field of 4.9 G, oriented vertically, allows for spectroscopic addressing of either $m_F$ state while on-site interactions enable spectroscopic addressing of either singly or multiply occupied states.
Atoms in singly occupied sites of a particular spin state, $|g\rangle = |{}^1\mathrm{S}_0, m_F\rangle$ for $m_F=+1/2$ or $-1/2$, are transferred to the long-lived excited state $|e\rangle = |{}^3\mathrm{P}_0, m_F\rangle$  using a $\pi$ pulse from an ultrastable clock laser (26 mHz linewidth) \cite{Bishof2013}. We remove any remaining atoms in other spin states or in multiply occupied sites. 
We then interrogate the $e$ atoms using Ramsey spectroscopy \cite{Ramsey1950}, first by placing atoms in a superposition state $|g\rangle + |e\rangle$ with a $\pi/2$ pulse, then waiting several seconds as the two states acquire a relative phase shift $\phi$, with a state $|g\rangle + e^{-i \phi}|e\rangle$.
A final $\pi/2$ pulse converts this phase difference into a population difference, $i\sin \phi |g\rangle + \cos \phi |e\rangle$.

We use state-dependent absorption imaging to measure the spatial distribution of the $|g\rangle$ and $|e\rangle$ state populations in the horizontal plane, from which we infer the atomic clock frequency distribution \cite{Marti2014}.
The $g$ atoms are imaged with a $5$ or $10\;\mu$s pulse of resonant 461 nm light and subsequently removed. The $e$ atoms are then repumped to $g$ and imaged with a second pulse of resonant light \cite{ImagingSupplemental}.
A final pulse without atoms is used to acquire a reference image.
The data are processed to generate column densities of the ground state $\tilde n_g$ (Fig.~\ref{fig:schematic}b), excited state $\tilde n_e$ (Fig.~\ref{fig:schematic}c), and normalized excitation fraction $\tilde p_e = \tilde n_e / (\tilde n_e + \tilde n_g)$ (Fig.~\ref{fig:schematic}d).
Imaging is done at saturation intensity for the best signal-to-noise ratio \cite{Reinaudi2007, Maussang2010, Muessel2013, Marti2016a}.

Spatial correlations of the density characterize an imaging system's resolution \cite{Hung2011b}.
Here, we measure correlations in the excitation fraction by placing atoms in the state $|g\rangle{+}|e\rangle$.
An imaging sequence projects the atomic wave function on each lattice site onto either $|g\rangle$ or $|e \rangle$. This projection produces well-calibrated binomial noise with zero correlation length.
The finite imaging resolution creates spatial correlations in the images (but not in the actual sample).
The measured spatial autocorrelation function (black points, Fig.~\ref{fig:schematic}), $\langle (\tilde p_e^\mathbf{i} - \bar p_e) (\tilde p_e^{\mathbf{i}+\mathbf{j}} - \bar p_e) \rangle / \mathrm{var}\;\tilde p_e$, corresponds to a $1/e^2$ radius imaging resolution of $1.1\;\mu\mathrm{m}$ for a $5\;\mu\mathrm{s}$ imaging pulse time (red line). 
Here, $\bar p_e$ is the average excitation fraction taken over all pixels $\mathbf{i}$. Longer pulse times have a slightly worse imaging resolution as atoms are accelerated out of the depth of field \cite{ImagingSupplemental}.

We use a series of images similar to Fig.~\ref{fig:schematic}d to determine small differences in the clock frequency across the lattice. 
Frequency shifts are measured by comparing the excitation fractions in one region against another, with a frequency uncertainty set by QPN, removing the laser frequency (or phase) noise.
Magnetic fields, interactions, or the lattice light can shift the local clock frequency.
A spatially varying clock transition frequency creates a spatially inhomogeneous excitation fraction,
\begin{equation}\label{eqn:excitation}
\tilde p_e(\mathbf{r}) = \frac{1}{2} + \frac{C}{2} \cos \left(2 \pi f(\mathbf{r}) T + \phi_0 \right),
\end{equation}
where $C$ is the contrast, $f(\mathbf{r})$ it the local clock frequency, $T$ is the Ramsey free-evolution time, and $\phi_0$ is a common-mode phase offset.
Small misalignments in the lattice beams and birefringence of the vacuum chamber windows induce a small gradient of the vector ac Stark shift.
A typical value is approximately $m_F\times20$ mHz across the 50 $\mu$m sample size, corresponding to a fictitious magnetic field gradient of $0.04\;\mathrm{G/cm}$.
Ordinarily, such a small gradient would be negligible in state-of-the-art optical lattice clocks, as this frequency shift is sufficiently small and can be averaged away using opposite $m_F$ states.
We apply an additional (real) magnetic field gradient, enabling us to either cancel or increase the overall spatial frequency shifts.
A parametric plot of the excitation fraction $P_1$ of region 1 against $P_2$ of region 2 (regions marked in Fig.~\ref{fig:signal_to_noise}a) shows a clear ellipse (Fig.~\ref{fig:signal_to_noise}b).
The eccentricity of the ellipse increases as the phase difference $2\pi (f_1 - f_2) T$ increases with longer interrogation times, where $f_i$ is the average frequency in region $i$.
For short times, $T=0.1$~s, we observe nearly perfect contrast in the excitation fraction with a small phase difference.
At longer times, the frequency difference becomes more apparent as an increasing eccentricity of the ellipse (Fig.~\ref{fig:signal_to_noise}b).
The contrast decreases with a time constant of 8 s, likely due to Raman scattering of lattice photons by $e$ atoms \cite{Hutson2017}. 

The competition between a linearly increasing phase shift and an exponentially decreasing contrast creates a maximum signal-to-noise at 4 s (Fig.~\ref{fig:signal_to_noise}c) \cite{ImagingSupplemental}.
The fractional uncertainty $\delta(f_1-f_2)/f_0$, is measured from $T=0.1$ to $20$  s (Fig.~\ref{fig:signal_to_noise}c, blue circles) and matches well with the calculated QPN limit of the 3,000 atoms in each rectangle (blue curve) for a fixed per cycle dead time of 16 s and total measurement time of 900 s.

This synchronous measurement technique can reach record-breaking performance for extended measurement times.
The fractional uncertainty from averaging 1,000 experimental repetitions over six hours (red square, Fig.~\ref{fig:signal_to_noise}c) is set by the QPN limit (red curve) and has an Allan deviation consistent with white noise (Fig.~\ref{fig:signal_to_noise}c, inset).
The uncertainty of $2.5\times 10^{-19}$, the best measured in any system, corresponds to a 100 $\mu$Hz frequency uncertainty, or $2.7\times10^{-3}$ rad uncertainty of a $1.08\times 10^{16}$ rad total phase shift over a 4 s coherent evolution time.

\begin{figure}[t!]
\begin{center}
\includegraphics[scale=1.0]{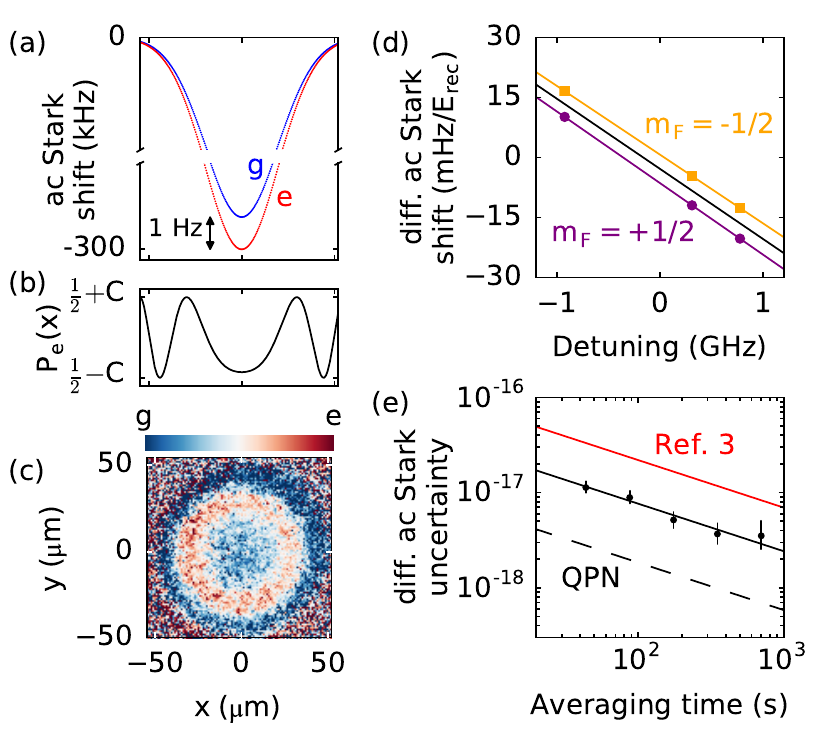}
\caption{Determining the lattice magic frequency through imaging spectroscopy.
(a) When the lattice frequency is higher than the magic frequency, $e$ atoms (red line) experience a deeper potential than $g$ atoms (blue line, not to scale).
(b) The mismatch in lattice potentials results in a spatially dependent clock frequency that we measure via $\tilde{p}_e$ after a Ramsey sequence, with a population that varies from $\frac{1}{2}+C$ to $\frac{1}{2}-C$, where $C$ is the contrast.
An image of $\tilde{p}_e$ from a single cycle of the experiment is shown in (c).
(d) A fit of the distance between the imaged rings gives a single-shot estimate of the differential ac Stark shift, which we determine separately for the $m_F{=}1/2$ (purple circles) and $-1/2$ (orange squares) states.
Detuning of the lattice light is from the scalar magic frequency 368.554 725 THz \cite{Shi2015}.
(e) The uncertainty in the differential ac Stark shift averages down rapidly (black circles, fit is black line), with a QPN limit (dashed black line) ten times better than the state-of-the-art \cite{Nicholson2015}.
\label{fig:ac_stark}}
\end{center}
\end{figure}

We use the spatial mapping of the clock frequency to rapidly determine the differential ac Stark shift induced by the optical trap. 
The uncertainty in this shift remains the second largest systematic effect in state-of-the-art clocks \cite{Nicholson2015}. 
As measurement precision improves, the higher-order effects in the $10^{-19}$ region will be investigated \cite{Brown2017}.
In previous work, measurements of the differential ac Stark shift were performed by asynchronously comparing the clock frequency in different lattice depths against an ultrastable cavity acting as a frequency flywheel.
These measurements are typically dominated by laser noise, yielding a precision worse than what could be achieved in a QPN-limited system \cite{Nicholson2015}.

Imaging spectroscopy allows for a measurement of the differential ac Stark shift within a single image.
In previous work, combining imaging with spectroscopy or interferometry has been used to measure the spatial distribution of scalar \cite{Bertoldi2010} and vector \cite{Vengalattore2007} ac Stark shifts, dipolar magnetic fields \cite{Marti2014}, and microwave field strength \cite{Boehi2010, Horsley2013}.
Here, we apply imaging spectroscopy to evaluate the differential ac Stark shift in a record short time.
The vertical lattice beam, which propagates along the imaging axis, is used to create a spatially inhomogeneous clock frequency because the optical lattice intensity varies with the Gaussian profile of the trapping laser beam (Fig.~\ref{fig:ac_stark}a).
The \textit{local} clock frequency $f(\mathbf{r})$ then varies spatially, depending on the \textit{local} lattice intensity $I(\mathbf{r})$.
We measure the position-dependent excitation fraction as described in Eq.~\ref{eqn:excitation} and shown schematically in Fig.~\ref{fig:ac_stark}b.
A sample image (Fig.~\ref{fig:ac_stark}c) of the excitation fraction changes radially from the center of the trap as the optical lattice intensity decreases.
We exaggerate the effect by detuning the frequency of the vertical lattice beam by 1 GHz from the magic frequency, the frequency of the laser at which the differential ac Stark shift vanishes.
This large detuning creates an overall 1 Hz mismatch of the $|g\rangle$ and $|e\rangle$ state potentials, out of a total ac Stark shift of 300 kHz (Fig.~\ref{fig:ac_stark}a).
We choose $T=4$ s, such that in each realization, atoms on the edge of the trap wrap greater than a $2\pi$ rad phase shift as compared to the center (Fig.~\ref{fig:ac_stark}b), leading to clear rings in the local excitation fraction (Fig.~\ref{fig:ac_stark}c).

Each image is fit to a model including a local frequency shift proportional to the local intensity $f(\mathbf{r}) = f_0 + a(f_\mathrm{lattice}) I(\mathbf{r})$, where $I(\mathbf{r})$ is local intensity of the vertical lattice beam and $a(f_\mathrm{lattice})$ is our fit determining the differential ac Stark shift. The intensity of the vertical optical lattice beam is measured with motional sideband spectroscopy \cite{Blatt2009}, while the beam waist is determined independently.
We measure the coefficient $a(f_\mathrm{lattice})$ at three trapping light frequencies for both $m_F=\pm 1/2$ spin states (Fig.~\ref{fig:ac_stark}d).
Averaging the coefficient for the two spin states removes the vector ac Stark shift and allows us to determine the combined scalar plus tensor differential ac Stark shift (black line, Fig.~\ref{fig:ac_stark}d) \cite{Campbell2017}.

\begin{figure}[t!]
\begin{center}
\includegraphics[scale=1.0]{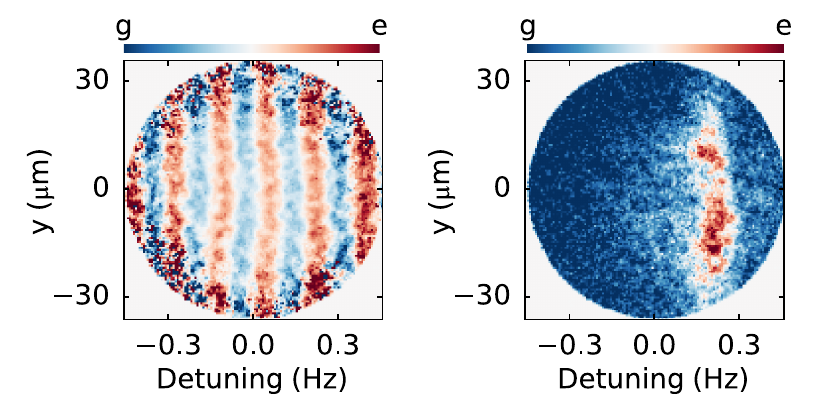}
\caption{Synchronous Rabi and Ramsey spectroscopy of an ultrastable laser. An applied magnetic field gradient creates a spatially dependent clock frequency.
This effectively `disperses' the signal by converting the frequency response into a spatial pattern, a multiplexed measurement of the laser noise and atomic response.
(a) In Ramsey spectroscopy, nearly all laser noise is common mode. `Ramsey fringes' are imaged as a spatially sinusoidal excitation fraction that measures the linearly increasing frequency shift across the sample, with $T=6$ s.
(b) In Rabi spectroscopy, the entire lattice is illuminated by a clock pulse for 8 s. Only atoms in a narrow spatial region can be excited. Here, the measured linewidth is 150 mHz.
\label{fig:rabi}}
\end{center}
\end{figure}

Synchronous interrogation removes laser noise that limited previous asynchronous measurements.
The differential ac Stark shift uncertainty is limited by the statistical noise of the measured coefficient, with a standard error of the mean $\delta a(f_\mathrm{lattice})$.
The peak differential ac Stark shift uncertainty while operating the lattice at the measured magic frequency is then $U_\mathrm{op}\delta a(f_\mathrm{lattice})$, where $U_\mathrm{op}=30\;E_\mathrm{rec}$ is the operational lattice depth, $E_\mathrm{rec}=h\times 3.47\,\mathrm{kHz}$ is the lattice photon recoil energy and $h$ is the Planck constant.
In Ref.~\onlinecite{Nicholson2015}, laser noise dominated the measurement imprecision, requiring ${\sim}13$ hours of averaging to reach an uncertainty of $1.1\times 10^{-18}$.
Our measurement (black points) rejects laser noise and requires only one hour to reproduce this uncertainty.
We have not yet reached the ultimate limit of this technique, the QPN limit, which would require only 6 minutes of averaging.
We believe the measurement noise is in excess of QPN noise because this measurement is made with the entire sample, including the relatively low-density edge of the sample.
These rapid measurement times open the possibility of reaching an uncertainty at the $10^{-19}$ level and studying hyperpolarizability shifts \cite{Westergaard2011, Brown2017}.

Finally, we use the atomic sample in a highly dispersive, multiplexed measurement of the frequency noise of an ultrastable laser.
A magnetic field gradient 0.26 G/cm shifts the clock frequency by approximately 1 Hz across the sample. 
This, in combination with our 1.1 $\mu$m imaging resolution, yields a 14 mHz frequency resolution.
In this experiment, we use the $m_F=+9/2$ state for increased frequency sensitivity and calibrate the magnetic field gradient with Ramsey spectroscopy.
The magnetic field gradient converts Ramsey fringes, the sinusoidal response of atoms to a laser frequency, into a sinusoidal spatial response (Fig.~\ref{fig:rabi}a).
The range of the magnetic fields is sufficiently small, such that all atoms respond to the short laser pulses used in Ramsey spectroscopy.
Laser noise manifests itself as a shot-to-shot variation in the location of the spatial nodes, but it cannot be observed from a single image.
To measure the laser noise, we perform Rabi spectroscopy for 8 s, such that only atoms in a narrow spatial region, 8 $\mu$m width for the 100 mHz Fourier limit, can be excited by this laser.
Frequency excursions excite a spatially broadened region. In this way, imaging spectroscopy serves as a multiplexed measurement of laser frequency noise~\cite{Bishof2013}, and as a dispersive element that converts frequency shifts into position variations of the atomic excitation.
The observed linewidth of 150 mHz (Fig.~\ref{fig:rabi}), the narrowest Rabi line shape measured on an optical transition, is limited mainly by the 8 s $|e\rangle$ state lifetime in the lattice.

In conclusion, imaging spectroscopy on the clock transition of lattice-trapped degenerate AE fermions combines the use of micron-resolution spatial imaging with submillihertz frequency resolution for advancing the measurement capabilities of atomic clocks. Comparing the frequency shifts of separated regions within a 3D lattice allows for rapid and precise measurements, reaching a record frequency precision of $2.5\times 10^{-19}$ in 6 hours, limited by the QPN of 3,000 atoms interrogated coherently for 4 s.  We apply these techniques to determine the differential ac Stark shift more rapidly than all previous work. We also observe the narrowest atomic linewidth on an optical transition. High spatial and frequency resolution imaging spectroscopy will be used to explore many-body physics through maps of local position and frequency density of states, and test the interplay of quantum mechanics and general relativity on the millimeter scale.

\begin{acknowledgments}
We acknowledge technical contributions from W. Milner, E. Oelker, J. Robinson, L. Sonderhouse, and W. Zhang, and useful discussions with T. Bothwell, S. Bromley, C. Kennedy, D. Kedar, and S. Kolkowitz. This work is supported by NIST, DARPA, AFOSR-MURI, and Grant No. NSF-1734006. G.E.M. is supported by a postdoctoral fellowship from the National Research Council and A.G. is supported by a postdoctoral fellowship from the Japan Society for the Promotion of Science.
\end{acknowledgments}

\bibliographystyle{myprl}
\bibliography{main}

\clearpage

\ifshowsupplemental
\newpage
\appendix
\newcommand{\var}{\operatorname{var}}

\renewcommand{\theequation}{S\arabic{equation}}
\renewcommand{\thefigure}{S\arabic{figure}}

\onecolumngrid

\begin{center}
\Large{\bf{Supplemental Material}}
\end{center}

\section{Absorption Imaging}
During absorption imaging, atoms are illuminated with a $\tau=5$ or $10\;\mu\mathrm{s}$ long pulse of resonant 461 nm light at approximately saturation intensity. Atoms projected onto the $|g\rangle$ state will scatter, on average, $N_\mathrm{scatt}=\Gamma \tau/4 \approx 500$ photons for $\tau=10\;\mu\mathrm{s}$, where $\Gamma=2\pi\times 30\;\mathrm{MHz}$. Atoms projected onto the $|e\rangle$ state do not scatter photons. After the image is acquired, $e$ atoms are repumped on the ${}^3\mathrm{P}_0 \to {}^3\mathrm{S}_1$ transition, where they cascade to the ground state through ${}^3\mathrm{S}_1 \to {}^3\mathrm{P}_1 \to {}^1\mathrm{S}_0$. Some atoms may decay to the metastable ${}^3\mathrm{P}_2$ state, and so a second repump laser simultaneously drives the ${}^3\mathrm{P}_2\to {}^3\mathrm{S}_1$ transition.

Atoms undergo diffusive motion during imaging, moving in the transverse plane as they acquire random momentum kicks from photon scattering. The spread in the transverse position is approximately $(v_\mathrm{rec}/6) \sqrt{\Gamma \tau^3}=0.7\;\mu\mathrm{m}$ for $\tau=10\;\mu\mathrm{s}$, smaller than our imaging resolution (the scattering rate is $\Gamma/4$ for imaging at saturation intensity).

The most substantial blurring occurs when atoms are pushed out of the depth of field of the imaging system. During imaging, atoms experience an acceleration $a=\hbar k \Gamma/4m=5\times 10^5\;\mathrm{m\,s^{-2}}$, where $k=2\pi/(461\;\mathrm{nm})$ is the imaging light wavenumber and $m$ is the atomic mass. After approximately 30 scattering events, atoms are heated sufficiently such that they are no longer bound by the lattice. When $\tau=10\;\mu\mathrm{s}$, the atoms are accelerated a distance of $a \tau^2/2 = 23\;\mu$m, greater than the Rayleigh range $z_R=\pi w_0^2/\lambda = 6\;\mu\mathrm{m}$ expected for the design $\mathrm{NA}=0.23$. The blurring can be minimized by slightly defocusing the imaging system, allowing the atoms to accelerate through the focus. This effect limits the imaging resolution to a $1/e^2$ radius of $1.8\;\mu\mathrm{m}$, larger than the $1.0\;\mu\mathrm{m}$ diffraction limit. For $\tau=5\;\mu\mathrm{s}$, this effect  is negligible and the resolution should be limited by the imaging resolution. Our measured resolution of $1.1\;\mu\mathrm{m}$ for $\tau=5\;\mu\mathrm{s}$ is only ten percent worse than the design resolution. 

The acceleration can also detune atoms by increasing their Doppler shift.   At the end of the pulse, the overall Doppler shift is $ka\tau/2\pi=10\;\mathrm{MHz}$ for $\tau=10\;\mu\mathrm{s}$. This effect is small compared to the atomic linewidth and we can safely ignore it.

\section{Quantum-projection-noise limit of ellipse fitting}
What is the quantum projection noise (QPN) limit in measuring a frequency difference between two ensembles of $N_a$ atoms? Two regions ($1$ and $2$) with clock frequencies $\omega_1$ and $\omega_2$ are interrogated by a laser with frequency $\omega_L$. After a Ramsey pulse sequence, the excitation fractions in the two regions are,
\begin{align}
P_1 &= \frac{1}{2} \left( 1 + C \cos (\omega_1 - \omega_L) t \right) \\
P_2 &= \frac{1}{2} \left( 1 + C \cos (\omega_2 - \omega_L) t \right),
\end{align}
where $C$ is the contrast. It is easiest to rewrite this problem with two angles: the accumulated phase between atoms in the first region and the laser $\theta = (\omega_1 - \omega_L)t$ and the phase difference between the atoms in the two regions $\phi = (\omega_2 - \omega_1) t$. In our experiment, the phase noise of the laser is much greater that the phase noise of the atoms. The goal is to then extract information about $\phi$ even when the uncertainty in $\theta$ is substantial. Here, we derive the QPN limit of $\phi$ when $\theta$ is unknown, as would occur for measurement times longer than the laser's coherence time.

Ellipse fitting allows for the determination of $\phi$ when $\theta$ is unknown and uncontrolled (Fig. 2b). For convenience, we define $x = P_1 - \frac{1}{2}$ and $y = P_2 - \frac{1}{2}$ so that,

\begin{align}
x &=  \frac{C}{2} \cos \theta \nonumber\\
y &=  \frac{C}{2}  \cos (\theta + \phi). \label{eqn:xy}
\end{align}

The variables $x$ and $y$ have uncertainty because of the QPN of $N_a$ atoms in each region. The variance of $x$ is,
\begin{equation}\label{eqn:qpn}
\var x = \frac{1}{N_a} \left( \frac{1}{2} - x\right)\left( \frac{1}{2} + x \right),
\end{equation}
and a similar equation holds for the variance of $y$.  The noise in determining $\phi$ can be found from Eq.~\ref{eqn:xy},
\begin{equation}\label{eqn:varphi1}
\var \phi = \left| \frac{\partial \phi}{\partial x} \right|^2 \var x + \left| \frac{\partial \phi}{\partial y} \right|^2 \var y .
\end{equation}

This assumes no prior knowledge of $\theta$ and that $x$ and $y$ are distributed normally, a good approximation for large $N$.  The partial derivatives $\partial \{\phi, \theta \}/\partial \{x, y \}$ can be calculated from the inverse of the Jacobian, $\partial \{ x, y\} /\partial \{ \theta, \phi \}$,
\begin{equation}
J^{-1} =
\left( \begin{array}{cc}
\partial x/\partial \theta & \partial x/\partial \phi \\
\partial y/\partial \theta & \partial y/\partial \phi
\end{array} \right)
= \frac{C}{2} \left( \begin{array}{cc}
-\sin \theta & 0 \\
-\sin (\theta+\phi) & -\sin (\theta+\phi)
\end{array} \right).
\end{equation}

Inverting this matrix gives the desired derivatives.
\begin{equation}
J^{} =
\left( \begin{array}{cc}
\partial \theta/\partial x & \partial \theta/\partial y \\
\partial \phi/\partial x & \partial \phi/\partial y
\end{array} \right)
= \frac{2}{C} \left( \begin{array}{cc}
-\csc \theta & 0 \\
\csc \theta & -\csc (\theta+\phi)
\end{array} \right)
\end{equation}

From this we evaluate Eq.~\ref{eqn:varphi1},
\begin{equation}\label{eqn:varphi2}
\var \phi = \frac{4}{C^2} \left( \csc^2 \theta \var x + \csc^2 (\theta+\phi) \var y \right).
\end{equation}

When the measurement time exceeds the laser coherence time, $\theta$ will be random and uniformly distributed from $0$ to $2\pi$. In this experiment, we intentionally randomize $\theta$ to produce this uniform distribution. We should then average $\var \phi$ over the distribution of possible values of $\theta$. For $n$ measurements $\phi_1, \phi_2, \hdots$ of the (true) phases $\theta_1, \theta_2,\hdots$, the variance in $\phi$ is, 

\begin{equation}\label{eqn:thingToSolve}
\langle \var\phi \rangle_\theta = \left( \sum_i \frac{1}{\var \phi_i} \right)^{-1} = \frac{4}{C^2} \left( \int_0^{2\pi} \frac{d\theta}{2\pi} \frac{1}{\csc^2 \theta \; \var x + \csc^2 (\theta+\phi) \; \var y} \right)^{-1} \end{equation}

In the case where $C=1$ and $\phi=\pi/2$, when the parametric plot of $x$ vs $y$ traces a circle, then $\var x = (1 - \cos^2\theta)/4N_a$ and $\var y = (1 - \sin^2\theta)/4N_a$ (Eq.~\ref{eqn:qpn}). 

\begin{equation}\label{eqn:phiUnityContrast}
\langle \var \phi \rangle_\theta = 4 \left( \int_0^{2\pi} \frac{d\theta}{2\pi} 
\frac{1}{\csc^2 \theta \var x + \sec^2 \theta \var y} \right)^{-1}
= \frac{2}{N_a} \qquad \textrm{for }C=1
\end{equation}
This is the same as the result for a QPN-limited Ramsey spectroscopy. Naively, one might expect that the variance increases as $C^{-2}$. Numerically solving Eq.~\ref{eqn:thingToSolve} shows that the trend is slightly faster (Fig.~\ref{fig:qpn_contrast}). Counterintuitively, the QPN noise is maximized when $\phi=\pi/2$ (a circle on a parametric plot), and minimized for $\phi=0$ (a line). The fits are biased near $\phi\approx0$, and so we ensure that $\phi$ is large enough to avoid this problem.

\begin{figure}[!t]
\centering
\includegraphics[scale=1]{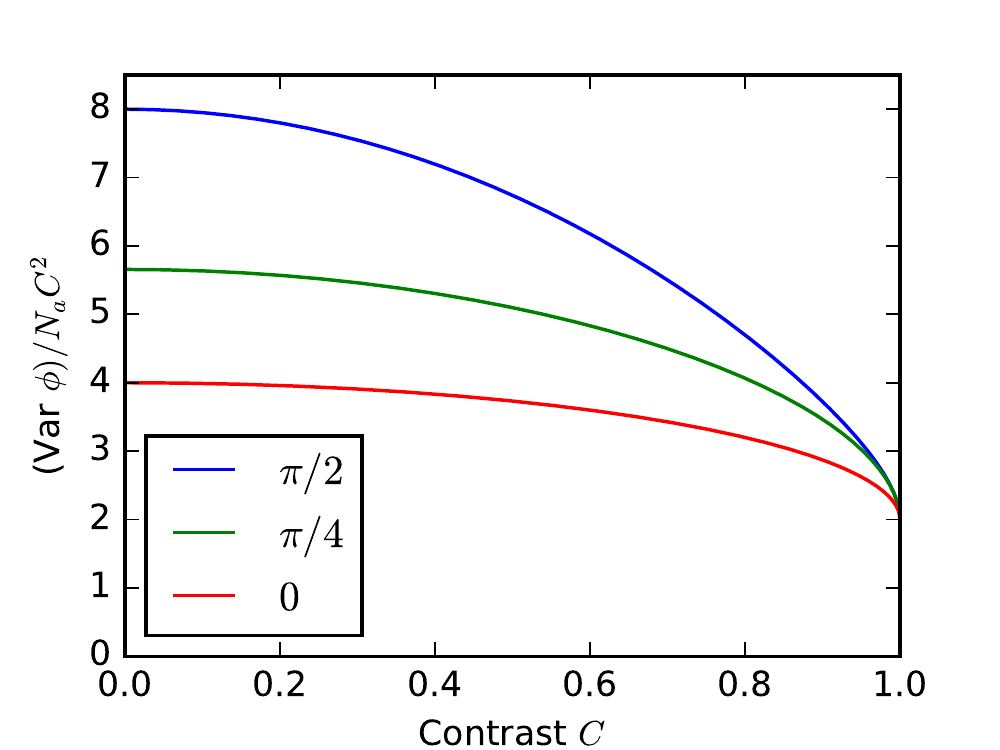}
\caption[something]{The variance of $\phi$ scales proportionally with $1/N_a C^2$. Plotted here is the proportionality constant, which increases as the contrast drops. Each curve corresponds to a different $\phi$ . While $\phi\approx\pi/2$ has the greatest QPN, it is often used because it typically has the lowest technical noise. For $C=1$, we have the usual QPN-limit for a single atom interferometer.\label{fig:qpn_contrast}}
\end{figure}

\subsection{Maximum Likelihood Estimator}

Many algorithms have been used to extract $\phi$ from parametric data \cite{Foster2002,Stockton2007a}. To generate a fit of $\phi$ that does not rely on the above model, we use a maximum likelihood estimator where the noise of $P_1$ and $P_2$ is a fit parameter. The overall in uncertainty in $\phi$ generated by this method might be slightly larger than the QPN as calculated in Eq.~\ref{eqn:qpn}. Experimentally, we observe that the increase in uncertainty is marginal and the results match well with theory (Fig.~2c).

\subsection{Jackknifing}

How do we calculate the uncertainty in the fits of $\phi$ from the maximum likelihood estimator, and verify that the uncertainty matches the QPN model (Eq.~\ref{eqn:thingToSolve} and Fig.~\ref{fig:qpn_contrast})? Most curve fitting algorithms that estimate the error of parameters in the model (e.g., $\phi$) require some prior knowledge of the distribution of the data, e.g. the noise of $P_1$ and $P_2$. Resampling techniques, such as jackknifing, are powerful methods that can estimate the uncertainty of parameters from the data alone, without any prior knowledge of the underlying distributions. 

We use the simplest form of jackknifing, removing a single datum and fitting the remaining data \cite{Miller1974}. This procedure is repeated over all permutations. Let $\bar\phi_{\neq i}^\mathrm{JK}$ be a fit to all data \textit{except} the $i^\mathrm{th}$ data point. The core concept is that the distribution of $\bar\phi_{\neq i}^\mathrm{JK}$ mimics the true distribution of  $\phi$. We can then approximate the variance of $\phi$ as,

\begin{equation}
\var \phi \approx \var \phi^\mathrm{JK} = \frac{n-1}{n} \sum \left( \bar \phi^\mathrm{JK} - \bar\phi_{\neq i}^\mathrm{JK} \right)^2,
\end{equation} 
where $\bar \phi^\mathrm{JK}$ is the mean of $\bar\phi_{\neq i}^\mathrm{JK}$ and $n$ is the total number of measurements. Estimates are shown in Fig. 2c.

An important application of jackknifing is to estimate the knowledge gained during each experimental cycle. To construct an Allan deviation from ellipse fitting, it is necessary to calculate a phase estimate $\phi^\mathrm{JK}_i$ for the $i^\mathrm{th}$ measurement. We accomplish this by measuring how much the $i^\mathrm{th}$  datum `pulls' the overall fit,
\begin{equation}
\phi_i^\mathrm{JK} = n\, \bar \phi^\mathrm{JK} - (n-1)\, \bar \phi^\mathrm{JK}_{\neq i} .
\end{equation}
From this we can calculate a series of single-shot estimates of the difference frequency $\omega_1^i - \omega_2^i = \phi^\mathrm{JK}_i/t$, where $t$ is the Ramsey dark time. The Allan deviation calculated from this method is plotted in the inset to Fig. 2c.

\fi

\end{document}